\documentclass[aps,prl,reprint,showpacs,preprintnumbers,nofootinbib,superscriptaddress]{revtex4-1}
\usepackage{slashed}
\usepackage{bm} 
\usepackage{epsf}
\usepackage{graphicx,epsfig}
\usepackage{latexsym,amssymb,amsmath,float,url}
\usepackage{bbold}
\usepackage{latexsym}
\usepackage{soul}
\usepackage{dsfont}

\input{epsf}

\begin{document}

\title{Particle-antiparticle asymmetries from annihilations}

\author{Iason\ Baldes}
\email{Corresponding author: ibaldes@student.unimelb.edu.au}
\affiliation{ARC Centre of Excellence for Particle Physics at the Terascale,
School of Physics, The University of Melbourne, Victoria 3010, Australia}

\author{Nicole F.\ Bell} 
\affiliation{ARC Centre of Excellence for Particle Physics at the Terascale,
School of Physics, The University of Melbourne, Victoria 3010, Australia}

\author{Kalliopi\ Petraki}
\affiliation{Nikhef, Science Park 105, 1098 XG Amsterdam, The Netherlands}

\author{Raymond R.\ Volkas}
\affiliation{ARC Centre of Excellence for Particle Physics at the Terascale,
School of Physics, The University of Melbourne, Victoria 3010, Australia}

\date{July 17, 2014}

\preprint{}

\begin{abstract}

An extensively studied mechanism to create particle-antiparticle
asymmetries is the out-of-equilibrium and CP violating decay of a
heavy particle.  Here we instead examine how asymmetries can arise
purely from $2 \to 2$ annihilations rather than from the usual $1 \to
2$ decays and inverse decays.  We review the general conditions on the
reaction rates that arise from $S$-matrix unitarity and CPT
invariance, and show how these are implemented in the context of a
simple toy model. We formulate the Boltzmann equations for
this model, and present an example solution.
\end{abstract}



\maketitle
\emph{Introduction.} --
Cosmological observations have shown $ \Omega_{DM} \approx 5\Omega_{B} \approx 0.24$, where $\Omega_{DM(B)}$ is the dark matter (baryon) density divided by the critical density~\cite{Hinshaw:2012aka,Ade:2013zuv}. However, current physics cannot explain what makes up $\Omega_{DM}$, why the baryon asymmetry of the universe (BAU) and hence $\Omega_{B}$ is non-negligible~\cite{PhysRevLett.17.712}, or indeed why $\Omega_{B} \sim \Omega_{DM}$. 
A baryogenesis mechanism satisfying the Sakharov conditions -- violation of the baryon number, violation of charge conjugation (C) and charge parity (CP) symmetries, and a departure from thermal equilibrium -- is required to explain the BAU~\cite{Sakharov:1967dj}. A similar asymmetry may also exist in the DM sector. In fact, asymmetric DM (ADM) scenarios seek to explain $\Omega_{B} \sim \Omega_{DM}$ as resulting from $n_{B}\sim |n_{X}-n_{\overline{X}}|$, where $n_{B}$ is the baryon number density and $n_{X} (n_{\overline{X}})$ is the DM particle (antiparticle) density~\cite{Nussinov:1985xr,Davoudiasl:2012uw,Petraki:2013wwa,Zurek:2013wia}. 
Understanding possible mechanisms for creating particle-antiparticle asymmetries is therefore crucial if we are to understand the cosmological history of the universe at the earliest times. 

In well known scenarios of baryogenesis, a matter-antimatter asymmetry
is created by the out-of-equilibrium decay of a heavy particle
\cite{Weinberg:1979bt,Kolb:1979qa,PhysRevLett.45.2074,Fukugita:1986hr}.
Similar mechanisms have been applied to ADM scenarios
\cite{Davoudiasl:2010am}.  The decays must be CP violating for a
preference of matter to be created over antimatter. Furthermore, the
asymmetry can only be created once the decaying particle has departed
from thermal equilibrium, because $S$-matrix unitarity ensures no net
preference for particle over antiparticle states can occur in
equilibrium.  Such scenarios have been studied
extensively.

In contrast there has been much less focus on asymmetries created from annihilations. Again, due to the unitarity, one or more of the particles involved in the annihilation must go out of thermal equilibrium for an asymmetry to be generated \cite{PhysRevLett.41.281,PhysRevLett.42.746,PhysRevD.19.3803}. This is the case in WIMPy baryogenesis, for example, in which heavy neutral particles freeze out and become the DM density and at the same time create the BAU through their annihilations \cite{Cui:2011ab,Bernal:2012gv,Bernal:2013bga,Kumar:2013uca,Racker:2014uga}. 
The effect of $2 \leftrightarrow 2$ annihilations has also been
investigated in the context of leptogenesis~\cite{Pilaftsis:2003gt,Pilaftsis:2005rv,Nardi:2007jp,Davidson:2008bu,Fong:2010bh}. In this case, it was found that the
annihilations change the asymmetry at high temperature but have only a
negligible effect on the final asymmetry~\cite{Pilaftsis:2003gt}.  However, there is no reason
to expect this feature to hold for baryogenesis in general.

The effect of annihilations is therefore interesting from -- at least -- the perspective of baryogenesis. The WIMPy baryogenesis mechanism also explains the DM density, but with no asymmetry between DM particles and antiparticles. However, it may be possible to construct an ADM model in which such annihilations play a role: this paper is a first step towards such a goal.\footnote{Such a model was constructed previously, however we find the unitarity constraint was not properly taken into account \cite{Farrar:2005zd}.}$^{,}$\footnote{Asymmetry creation during freeze-in has also been considered~\cite{Bento:2001rc,Hook:2011tk,Unwin:2014poa}. We are instead concerned with freeze-out.}

The purpose of this paper is to provide a general framework for models which seek to create particle-antiparticle asymmetries from annihilations. While certain aspects of such mechanisms are necessarily model dependent, other considerations, such as the unitarity relations and construction of the Boltzmann equations are generic. Our focus in this paper is on examining asymmetries from annihilations alone; in accompanying work we examine scenarios in which decays and annihilations compete in creating the final asymmetry~\cite{Baldes:2014rda}.

The structure of the paper is as follows. In the next section we review $S-$matrix unitarity and its implications for the CP violating reaction rates of annihilations. We then study a toy model involving the interaction between four fermions. We outline the Boltzmann equations for the model and show a non-zero source term develops when one or more of the species depart from equilibrium. We calculate the relevant thermally averaged cross sections and solve the Boltzmann equations numerically.

\emph{$S-$Matrix Unitarity and Time Reversal.} --
Unitarity of the $S$-matrix ($S^{\dagger}S=SS^{\dagger}=1$) together with invariance under charge parity time (CPT) implies for the usual invariant matrix elements:
	\begin{align}
	\sum_{\beta}|\mathcal{M}(\alpha \to \beta)|^{2} = \sum_{\beta}|\mathcal{M}(\beta \to \alpha)|^{2} \nonumber \\
	= \sum_{\beta}|\mathcal{M}(\overline{\beta} \to \overline{\alpha})|^{2} = \sum_{\beta}|\mathcal{M}(\overline{\alpha} \to \overline{\beta})|^{2},
	\end{align}
where $\alpha$ is an arbitrary state, $\overline{\alpha}$ its CP conjugate and the sum runs over all possible states $\beta$. Consider the collision term in the Boltzmann equations for the transition of a set of particles $\alpha_{i}$ where $i=1,...,n$ to and from a set of particles $\beta_{j}$ where $j=1,...,m$. Let us denote the integrated collision term for transitions $\alpha \to \beta$ in chemical equilibrium as $W(\alpha \to \beta)$. Approximated using Maxwell-Boltzmann statistics the net collision term is related to the matrix elements by \cite{earlyuniverse}:
	\begin{align}
	&W(\beta \to \alpha)-W(\alpha \to \beta) = \nonumber \\
	&\int...\int   d\Pi_{\alpha 1}...d\Pi_{\alpha n}d\Pi_{\beta 1}...d\Pi_{\beta m}\delta^{4}\left(\sum p_{i} - \sum p_{j}\right)(2\pi)^{4} \nonumber \\
		& \times\Big\{f_{\beta 1}...f_{\beta m}|\mathcal{M}(\beta \to \alpha)|^{2} -f_{\alpha 1}...f_{\alpha n}|\mathcal{M}(\alpha\to \beta)|^{2}\Big\}, \label{eq:maxboltcollision}
	\end{align}
where $f_{\psi}=\mathrm{Exp}[(\mu_{\psi}-E_{\psi})/T]$ is the phase space density of species $\psi$ with chemical potential $\mu_{\psi}$ at energy $E_{\psi}$,
	\begin{equation} 	
	d\Pi_{\psi}=\frac{g_{\psi}d^{3}p_{\psi}}{2E_{\psi}(2\pi)^{3}}
	\end{equation}
is the normalized volume element of the three momenta, $g_{\psi}$ are the degrees of freedom, and we assume throughout \emph{kinetic} equilibrium so that the temperature ($T$) of each species is identical. Under \emph{chemical} equilibrium we have in addition,
	\begin{equation}
	\sum_{i}\mu_{\alpha i} = \sum_{j}\mu_{\beta j}.
	\label{eq:chemeq}
	\end{equation}
Chemical equilibrium and the delta function enforcing four momentum conservation allows the replacement:
	\begin{equation}
	f_{\beta 1}...f_{\beta m} \to f_{\alpha 1}...f_{\alpha n},
	\label{eq:replace}
	\end{equation}
under the integral sign in Eq. (\ref{eq:maxboltcollision}). Using the replacement in Eq. (\ref{eq:replace}) and taking the sum over all possible final states one finds \cite{Dolgov:1979mz}:
	\begin{align}
	\label{eq:eqcond}
	\sum_{\beta}W(\alpha \to \beta) = \sum_{\beta}W(\beta \to \alpha) 
\nonumber\\
	= \sum_{\beta}W(\overline{\beta} \to \overline{\alpha}) = \sum_{\beta}W(\overline{\alpha} \to \overline{\beta}),
	\end{align}
where the second line follows from CPT invariance. Equation (\ref{eq:eqcond}) means there must be a departure from thermal equilibrium for a baryon asymmetry to be produced (the third Sakharov condition).\footnote{An exception is the spontaneous baryogenesis scenario, in which CPT is violated spontaneously by the expansion of the universe, but the particles themselves remain in thermal equilibrium \cite{Cohen:1987vi,Cohen:1988kt}.}$^{,}$\footnote{The same result holds for full quantum statistics. The collision term and phase space densities are modified to take into account quantum statistics \cite{earlyuniverse}, but the unitarity condition is also modified~\cite{Weinberg:1979bt,Kolb:1979qa,Hook:2011tk}.} We will apply this unitarity constraint below so as to correctly relate the CP violation in the reaction rates which enter the Boltzmann equations~\cite{Toussaint:1978br,Bhattacharya:2011sy}.

\emph{Toy model.} --
Consider the interaction Lagrangian:
	\begin{align}
	\mathcal{L} & = \frac{1}{4}\kappa_{1}\overline{\Psi_{1}^{c}}\Psi_{1}\overline{f^{c}}f+\frac{1}{4}\kappa_{2}\overline{\Psi_{2}^{c}}\Psi_{2}\overline{f^{c}}f+\frac{1}{2}\kappa_{3}\overline{\Psi_{2}^{c}}\Psi_{1}\overline{f^{c}}f \nonumber	\\
	&\;+\frac{1}{2}\lambda_{1}\overline{\Psi_{2}^{c}}\Psi_{1}\overline{\Psi_{1}}\Psi_{1}^{c}+\frac{1}{4}\lambda_{2}\overline{\Psi_{2}^{c}}\Psi_{2}\overline{\Psi_{1}}\Psi_{1}^{c}+\frac{1}{2}\lambda_{3}\overline{\Psi_{2}^{c}}\Psi_{2}\overline{\Psi_{2}}\Psi_{1}^{c} \nonumber \\ & \; + H.c. 
	\end{align}
where the $\Psi$ and $f$ are Dirac fermions and the $\kappa_{i}$ and $\lambda_{i}$ are effective couplings with mass dimension -2.

The above Lagrangian violates the particle numbers associated with $\Psi_{1}$, $\Psi_{2}$ and $f$ but preserves the linear combination $\Delta(\Psi_{1}+\Psi_{2}-f)$. We will show how these interactions  will generate an asymmetry in the $f$ sector and a related asymmetry in the $\Psi$ sector, $\Delta(f)=\Delta(\Psi_{1}+\Psi_{2})$, through $2 \leftrightarrow 2$ processes. The last three interaction terms break the particle numbers associated with $\Psi_{1}$ and $\Psi_{2}$ individually but preserve $\Delta(\Psi_{1}+\Psi_{2})$. These latter interactions must be included to allow CP violation to arise in the interference between tree and loop level diagrams. Majorana masses are prohibited by the global symmetry of the Lagrangian $\Delta(\Psi_{1}+\Psi_{2}-f)=0$. 

We assume $f$ are in thermal equilibrium with the radiation bath and that $\Psi_{1}$ and $\Psi_{2}$ are coupled to the radiation bath only through their interactions in the above Lagrangian. The asymmetries are generated  during the time when the $\Psi$ particles are going out-of-equilibrium. We take the $\Psi_{2}$ mass greater than the $\Psi_{1}$ mass ($M_{2} > M_{1}$) and also consider the decays of $\Psi_{2}$ below. 

\begin{figure}[t]
\begin{center}
\includegraphics[width=110pt]{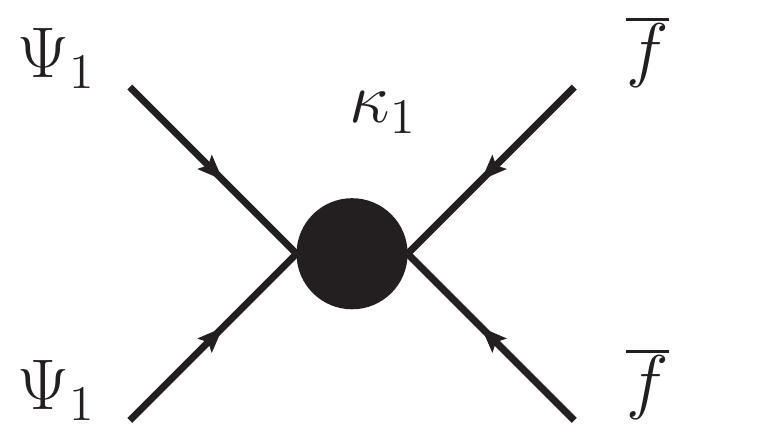}
\includegraphics[width=245pt]{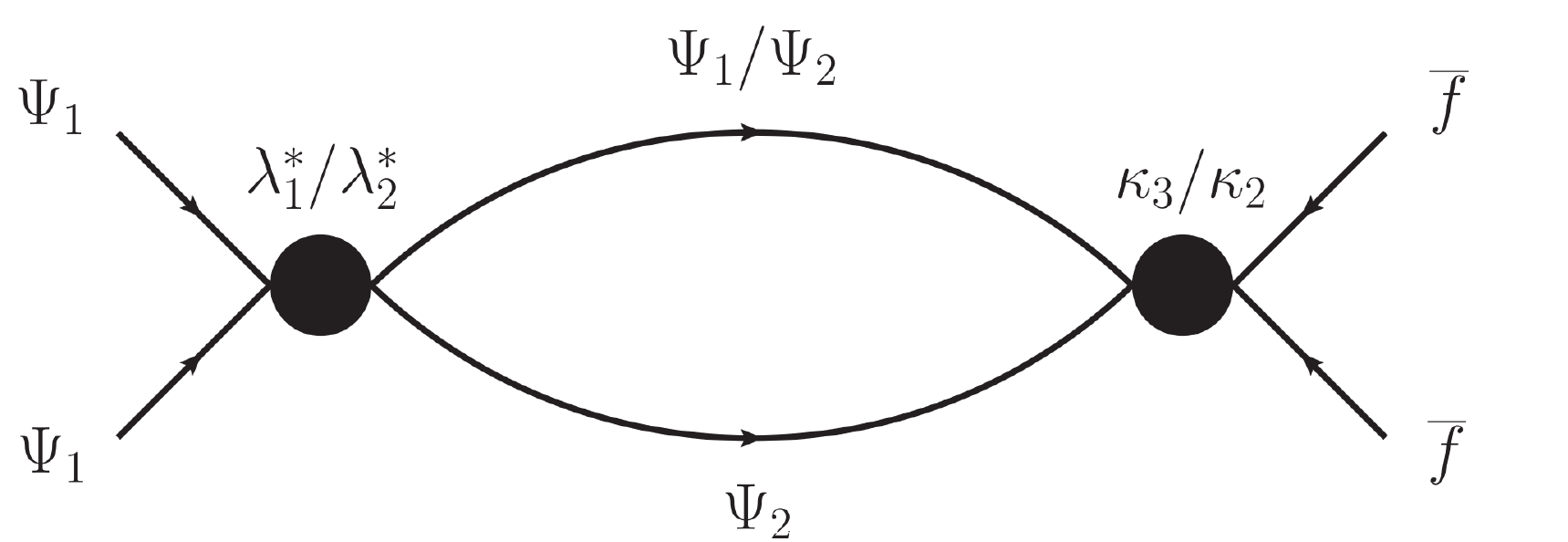}
\end{center}
\caption{Tree and one-loop diagrams for the annhilation $\Psi_{1}\Psi_{1}\to \overline{f}\overline{f}$.}
\label{fig:interference}
\end{figure}

The above Lagrangian includes four physical phases in the couplings. CP violation arises in $\Psi$ number changing interactions of the form $\Psi_{i}\Psi_{j} \to \overline{f}\overline{f}$ in the interference between the tree level and one loop level diagrams such as those depicted in Fig. \ref{fig:interference}.

We define the equilibrium reaction rate density -- which will enter as a collision term in the Boltzmann equation -- for the annihilation $\Psi_{1}\Psi_{1} \to \overline{f}\overline{f}$ as:
	\begin{align}
	(1+a_{1})A_{1} & \equiv W(\Psi_{1}\Psi_{1} \to \overline{f}\overline{f}) \\
		       &  = n_{\Psi 1}^{eq}n_{\Psi 1}^{eq}\langle v \sigma(\Psi_{1}\Psi_{1} \to \overline{f}\overline{f}) \rangle,
	\label{eq:a1}
	\end{align}
where the thermally averaged cross section comes from integrating over the phase space densities:
	\begin{align}
	n_{\alpha 1}^{eq}n_{\alpha 2}^{eq}\langle v \sigma(\alpha_{1}\alpha_{2} \to \beta_{1}\beta_{2}) \rangle &   \\
	\equiv \int...\int   d\Pi_{\alpha 1}d\Pi_{\alpha 2}d\Pi_{\beta 1} & d\Pi_{\beta 2}  \delta^{4}\left(\sum p_{i} - \sum p_{j}\right)(2\pi)^{4} \nonumber \\
		& \times f_{\alpha 1}^{eq}f_{\alpha 2}^{eq}|\mathcal{M}(\alpha_{1}\alpha_{2} \to \beta_{1}\beta_{2})|^{2}, \nonumber
	\end{align}
where $n_{\alpha i}^{eq}$ ($f_{\alpha i}^{eq}$) is the number (phase space) density in the absence of a chemical potential. We have parametrized the CP violation in the following way:
	\begin{align}
	a_{1} \equiv \frac{W(\Psi_{1}\Psi_{1} \to \overline{f}\overline{f})-W(\overline{\Psi_{1}}\overline{\Psi_{1}} \to ff)}{W(\Psi_{1}\Psi_{1} \to \overline{f}\overline{f})+W(\overline{\Psi_{1}}\overline{\Psi_{1}} \to ff)},
	\end{align}
hence the time reversed rate can be found by making the substitution: $a_{1} \to -a_{1}$. The other CP violating interactions are denoted:
\begin{align}
W(\Psi_{2}\Psi_{2} \to \overline{f}\overline{f}) \equiv (1+a_{2})A_{2}, \\
W(\Psi_{1}\Psi_{2} \to \overline{f}\overline{f})  \equiv (1+a_{3})A_{3}, \\
W(\Psi_{1}\Psi_{1} \to \Psi_{1}\Psi_{2}) \equiv (1+a_{4})A_{4},  \\
W(\Psi_{1}\Psi_{1} \to \Psi_{2}\Psi_{2}) \equiv (1+a_{5})A_{5},  \\
W(\Psi_{2}\Psi_{2} \to \Psi_{2}\Psi_{1}) \equiv (1+a_{6})A_{6}. 
\end{align}
CP conjugate rates can again be found by substituting $a_{i} \to -a_{i}$. The unitarity conditions yield:
\begin{align}
a_{1}A_{1}+a_{4}A_{4}+a_{5}A_{5}=0, \label{eq:unit1} \\
a_{2}A_{2}+a_{6}A_{6}-a_{5}A_{5}=0, \label{eq:unit2} \\
a_{3}A_{3}-a_{4}A_{4}-a_{6}A_{6}=0. \label{eq:unit3} 
\end{align}
We have checked that the CP violating rates calculated in terms of the underlying parameters of the Lagrangian do indeed respect these unitarity conditions. Note for $\kappa_{i}=\lambda_{i}\equiv \kappa$ the CP violation scales as $a_{i} \sim \kappa T^{2}$ for $T \gg M_{2}$ and $a_{i} \sim \kappa M_{2}^{2}/(8\pi)$ for $T \lesssim M_{2}$ except for $a_{1}$ which becomes kinematically suppressed at low $T$ (as $M_{2}>M_{1}$).

Washout interactions of the form $\Psi_{i}f \to \overline{\Psi_{j}}\overline{f}$ must also be taken into account. Furthermore sufficiently rapid interactions of the form $\overline{\Psi_{i}}\Psi_{j} \leftrightarrow \overline{\Psi_{k}}\Psi_{l}$ relate the chemical potentials of $\Psi_{1}$ and $\Psi_{2}$, these are also included in our numerical solutions below. These rates are denoted as:
	\begin{align}
	&W(\Psi_{1}f \to \overline{\Psi_{1}}\overline{f})=W_{1}, 
	&W(\Psi_{2}f \to \overline{\Psi_{2}}\overline{f})=W_{2}, \nonumber\\
	&W(\Psi_{1}f \to \overline{\Psi_{2}}\overline{f})=W_{3}, 
	&W(\Psi_{1}\overline{\Psi_{1}} \to \Psi_{2}\overline{\Psi_{1}}) = Z_{1}, \nonumber\\
	&W(\Psi_{1}\overline{\Psi_{2}} \to \Psi_{2}\overline{\Psi_{1}}) = Z_{2},
	&W(\Psi_{2}\overline{\Psi_{2}} \to \Psi_{1}\overline{\Psi_{2}}) = Z_{3}. \nonumber
	\end{align}
A priori $\Psi_{2}$ may have two decay channels:
	\begin{align}
	 \Gamma(\Psi_{2} \to \overline{\Psi_{1}}\overline{f}\overline{f}) = (1+\gamma_{a})\Gamma_{2a}, \\
	 \Gamma(\Psi_{2} \to \overline{\Psi_{1}}\Psi_{1}\Psi_{1}) = (1+\gamma_{b})\Gamma_{2b},
	\end{align}
where the $\gamma_{i}$ denote the CP odd component. Unitarity implies $\gamma_{a}\Gamma_{2a}=-\gamma_{b}\Gamma_{2b}$. Here we kinematically forbid the second decay channel, ensuring no CP violation is possible in the $\Psi_{2}$ decays. The remaining decay width is given by:
	\begin{equation}
	\Gamma_{2a} = \frac{|\kappa_{3}|^{2}(M_{2})^5}{3072\pi^3},
	\end{equation}
where we have ignored the final state masses. (We include the final state masses and the Lorentz factor suppression resulting from the thermal average in our numerical solutions.) 

\emph{Boltzmann equations.} --
We can now write down the Boltzmann equations using the usual approximation of Maxwell-Boltzmann statistics. The use of Maxwell-Boltzmann statistics allows one to factor out the chemical potential of a species from the collision term. The nonequilibrium rate is then simply the equilibrium rate multiplied by the ratio of the number density to the equilibrium number density of the incoming particles. For notational clarity we define the ratio of the number density to the equilibrium number density as:
	\begin{equation}
	r_{i} \equiv \frac{n_{i}}{n_{i}^{eq}}, \quad \quad \quad \quad \overline{r_{i}} \equiv \frac{n_{\overline{i}}}{n_{i}^{eq}}.
	\end{equation}
We assume $f$ and $\overline{f}$ are in thermal equilibrium with the radiation bath so $\mu_{f}=-\mu_{\overline{f}}$. We find the Boltzmann equations for $n_{1}$, $n_{2}$, and the asymmetries $n_{\Delta 1} \equiv n_{1}-n_{\overline{1}}$ and $n_{\Delta 2} \equiv n_{2}-n_{\overline{2}}$ in terms of the CP even and odd interaction rates. This results in a system of four coupled first order ordinary differential equations. The equations take the form:
	\begin{align}
	\frac{dn}{dt} +3Hn = (\text{source terms})+(\text{washout terms}),
	\end{align}
where $H$ is the Hubble rate, the source terms can create an asymmetry once one or more species depart from equilibrium and the washout terms drive towards equilibrium and washout any asymmetries present. For example, the equation for $n_{\Delta 1}$ has washout terms:
\begin{align}
&n_{2}^{eq}\Gamma_{2a}\Big[\overline{r_{2}}-r_{2}+\overline{r_{1}r_{f}r_{f}}-r_{1}r_{f}r_{f} \Big] \nonumber \\
&+2W_{1}\Big[\overline{r_{1}}\overline{r_{f}}-r_{1}r_{f}\Big] + W_{3}\Big[\overline{r_{2}}\overline{r_{f}}-r_{2}r_{f}+\overline{r_{1}}\overline{r_{f}}-r_{1}r_{f} \Big] \nonumber \\
&+Z_{1}\Big[r_{2}\overline{r_{1}}-\overline{r_{2}}r_{1}\Big]+2Z_{2}\Big[r_{2}\overline{r_{1}} - r_{1}\overline{r_{2}} \Big] + Z_{3} \Big[ r_{2}\overline{r_{1}}-r_{1}\overline{r_{2}} \Big] \nonumber \\
&+2A_{1} \Big[ \overline{r_{f}}\overline{r_{f}} - r_{f}r_{f} + \overline{r_{1}}\overline{r_{1}} - r_{1}r_{1} \Big] \nonumber \\
&+A_{3}\Big[ \overline{r_{f}}\overline{r_{f}}-r_{f}r_{f}+\overline{r_{2}r_{1}}-r_{2}r_{1} \Big] \nonumber \\
&+ A_{4} \Big[ r_{2}r_{1}-\overline{r_{2}r_{1}}+\overline{r_{1}r_{1}}-r_{1}r_{1} \Big] \nonumber \\
&+2A_{5} \Big[ r_{2}r_{2} - \overline{r_{2}r_{2}}+ \overline{r_{1}r_{1}} - r_{1}r_{1} \Big] \nonumber \\
& + A_{6} \Big[ r_{2}r_{2} -\overline{r_{2}r_{2}} +\overline{r_{1}r_{2}} -  r_{1}r_{2} \Big].
\end{align}
The source terms for $n_{\Delta 1}$ are:
	\begin{align}
&-2a_{1}A_{1}\Big[\overline{r_{f}r_{f}}+r_{f}r_{f}+\overline{r_{1}r_{1}}+r_{1}r_{1} \Big] \nonumber \\
&-a_{3}A_{3}\Big[\overline{r_{f}r_{f}}+r_{f}r_{f}+\overline{r_{2}r_{1}}+r_{2}r_{1} \Big] \nonumber \\
&-a_{4}A_{4}\Big[\overline{r_{1}r_{1}}+r_{1}r_{1}+\overline{r_{2}r_{1}}+r_{2}r_{1} \Big] \nonumber \\
&-2a_{5}A_{5}\Big[\overline{r_{2}r_{2}}+r_{2}r_{2}+\overline{r_{1}r_{1}}+r_{1}r_{1} \Big] \nonumber \\
&+a_{6}A_{6}\Big[\overline{r_{2}r_{2}}+r_{2}r_{2}+\overline{r_{2}r_{1}}+r_{2}r_{1} \Big].  
	\end{align}
By the application of the unitarity conditions (\ref{eq:unit1}-\ref{eq:unit3}) these terms can \emph{only} generate asymmetries, $n_{\Delta 1} \neq 0$, when the distribution of $\Psi$ particles depart from equilibrium: $r_{i} \neq 1$.

\begin{figure}[t]
\begin{center}
\includegraphics[width=245pt]{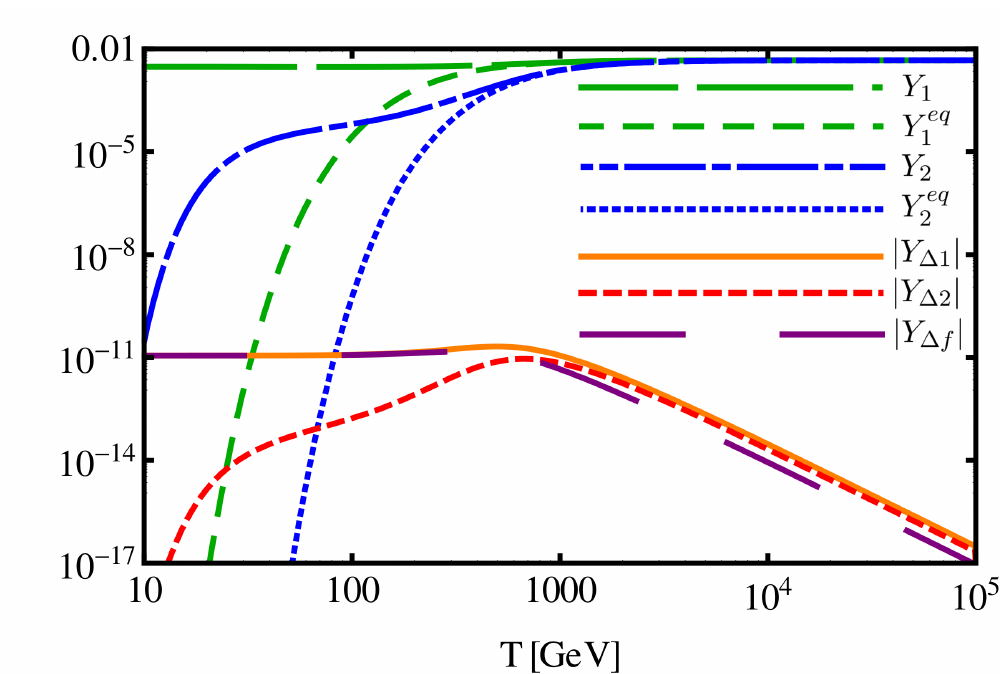}
\end{center}
\caption{Example solution to the system of coupled Boltzmann equations with densities normalized to the entropy density $Y_{\psi}\equiv n_{\psi}/s$ and shown evolving with temperature $T$, time proceeds right to left. Parameters are set to $M_{f}=100$ GeV, $M_{1}=800$ GeV, $M_{2}=2$ TeV, $|\kappa_{i}|=|\lambda_{i}|=5\times 10^{-13}$ GeV$^{-2}$, $\kappa_{3}=e^{-i3\pi/4}|\kappa_{3}|, \; \lambda_{1}=e^{i\pi/3}|\lambda_{1}|, \; \lambda_{2}=e^{-i\pi/6}|\lambda_{2}|, \; \lambda_{3}=e^{-i\pi/4}|\lambda_{3}|$.}
\label{fig:bmannsol}
\end{figure}

We proceed to solve the Boltzmann equations numerically. The standard change of variable is made to express the equations in terms temperature rather than time. We calculate the relevant cross sections and find the thermal averaged cross sections numerically by making use of the single integral formula \cite{Edsjo:1997bg}:
	\begin{align}
	&\langle v \sigma(ij \to \mathrm{final}) \rangle  \\
	&\quad = \frac{g_{i}g_{j}T}{8\pi^{4}n_{i}^{eq}n_{j}^{eq}} \int_{(m_{j}+m_{i})^{2}}^{\Lambda^{2}}p_{ij}E_{i}E_{j}v_{rel}\sigma K_{1}\left(\frac{\sqrt{s}}{T}\right) ds, \nonumber
	\end{align}
where $s$ is the centre-of-mass energy squared, $p_{ij}$ is the initial centre-of-mass momentum, $K_{1}(x)$ is the modified Bessel function of the second kind of order one and $\Lambda$ is the effective theory cut-off. Having calculated the reaction rates and CP violation, we then solve the system of coupled Boltzmann equations using Mathematica \cite{mathematica}. An example solution is shown in Fig.~\ref{fig:bmannsol}. 

The thermal history proceeds as follows. At high temperatures the $2 \leftrightarrow 2$ annihilations keep $\Psi_{1}$ and $\Psi_{2}$ close to thermal equilibrium and only a small asymmetry can develop (due to the expansion term the particles are never \emph{exactly} in equilibrium). The departure from equilibrium and hence the asymmetries increase as $T$ decreases and the reactions become less effective. At some point the $\Psi_{i}$ effectively decouple and the overall asymmetry remains constant. In Fig.~\ref{fig:bmannsol} this occurs around $T \approx 400$ GeV. Crucial to obtaining an asymmetry (with a common $T$ between sectors) is that at least some of the particles involved are massive: the decoupling of massless particles does not lead to $r_{i}\neq 0$. Numerically we find the maximum asymmetry is generated for decoupling at $T\sim M_{i}$. 

Eventually the heavier $\Psi_{2}$ decay into $\Psi_{1}$ and the final $\Delta(\Psi)$ asymmetry is stored in $\Psi_{1}$. Due to the different masses, couplings and phases, the asymmetries created in $\Psi_{2}$ and $\Psi_{1}$ are different and hence the eventual $\Delta(\Psi)$ decays of $\Psi_{2}$ do not washout the overall asymmetry.

Note that a large symmetric component of $\Psi_{1}$ is still present: $|Y_{\Delta 1}| \ll Y_{1}$. In a realistic model, so as to not overclose the universe, the symmetric component should be annihilated away. This can be achieved by introducing an interaction of the form $\overline{\Psi_{1}}\Psi_{1} \to \overline{f}f$. Alternatively $\Psi_{1}$ and $\overline{\Psi_{1}}$ could eventually decay. The asymmetry can then be stored in the decay products. These could be regular baryons or if they make up the DM, and have a sufficiently large annihilation cross section to annihilate away the symmetric component, form asymmetric DM \cite{Graesser:2011wi,Iminniyaz:2011yp,Lin:2011gj}.

We have assumed kinetic equilibrium for the $\Psi_{i}$ throughout. At high $T$ this is a good approximation as the $2 \leftrightarrow 2$ interactions effectively transfer momentum between the $\Psi_{i}$ and $f$. As we approach the decoupling point this approximation begins to breaks down~\cite{Hannestad:1999fj,Basboll:2006yx,Garayoa:2009my,HahnWoernle:2009qn}. This calculation can be further refined through the inclusion of departures from kinetic equilibrium, full quantum statistics and thermal masses which could give $\mathcal{O}(1)$ corrections to the final asymmetry.  

\emph{Conclusion.} --
We have presented a generic setup for the generation of
particle-antiparticle asymmetries from $2 \leftrightarrow 2$
processes, such as annihilations or scatterings.  This is to be
contrasted with the more well known scenario in which such asymmetries
are generated via $1 \rightarrow 2$ out-of-equilibrium decays.
We have explicitly outlined how the Boltzmann equations should be
formulated, taking $S$-matrix unitarity and CPT invariance into
account.  We have also presented an example numerical solution to the
Boltzmann equations in the context of a simple toy model.
Such techniques can be applied in calculation of particle-antiparticle
asymmetries in models of baryogenesis and ADM, as will be the focus of
our future work.

\smallskip

\emph{Acknowledgments.} -- IB was supported by the Commonwealth of Australia. NFB and RRV were supported in part by the Australian Research Council. KP was supported by the Netherlands Foundation for Fundamental Research of Matter (FOM) and the Netherlands Organisation for Scientific Research (NWO). IB would like to thank P. Cox and A. Millar for clarifying discussions. Feynman diagrams drawn using Jaxodraw \cite{Binosi:2003yf}.


%

\end{document}